\documentclass[12pt,preprint2]{aastex}
\usepackage{natbib}
\usepackage{color}
\usepackage{multirow}
\usepackage{ gensymb }

\shorttitle{Fomalhaut~b: Independent analysis}
\shortauthors{Galicher et al.}

\begin{document}

\title{Fomalhaut~b: Independent Analysis of the Hubble Space Telescope Public Archive Data}

\author{Rapha\"el Galicher\altaffilmark{1,2,3}, Christian Marois\altaffilmark{1}, B. Zuckerman\altaffilmark{4}, Bruce~Macintosh\altaffilmark{5}}
\affil{\altaffilmark{1}National Research Council Canada, Dominion Astrophysical Observatory, 5071 West Saanich Road, Victoria, BC, V9E 2E7, Canada;}
\affil{\altaffilmark{2}Dept. de Physique, Universit\'e de Montr\'eal, C.P. 6128 Succ. Centre-ville, Montr\'eal, Qc, H3C 3J7, Canada;}
\affil{\altaffilmark{3}LESIA, Observatoire de Paris, CNRS, UPMC, Universit\'{e} Paris Diderot, 5 place Jules Janssen, 92210 Meudon, France;}
\affil{\altaffilmark{4}Department of Physics and Astronomy, University of California, Los Angeles, CA
90095, USA;}
\affil{\altaffilmark{5}Lawrence Livermore National Laboratory, 7000 East Ave., Livermore, California, 94550, USA;}
    \email{raphael.galicher@obspm.fr}

\begin{abstract}
The nature and even the existence of a putative planet-mass companion ("Fomalhaut~b") to Fomalhaut has been debated since~2008.  In the present paper we reanalyze the multi-epoch ACS/STIS/WFC3 Hubble Space Telescope (HST) optical/near infrared images on which the discovery and some other claims were based.  We confirm that the~HST images do reveal an object in orbit around Fomalhaut but the detailed results from our analysis differ in some ways from previous discussions.  In particular, we do not confirm flux variability over a two-year interval at~$0.6\,\mu$m wavelength and we detect Fomalhaut~b for the first time at the short wavelength of~$0.43\,\mu$m. We find that the~HST image of Fomalhaut~b at~$0.8\,\mu$m may be extended beyond the~PSF. We cannot determine from our astrometry if Fomalhaut~b will cross or not the dust ring.  The optical through mid-infrared spectral energy distribution~(SED) of Fomalhaut~b cannot be explained as due to direct or scattered radiation from a massive planet. We consider two models to explain the~SED: (1) a large circumplanetary disk around an unseen planet and (2) the aftermath of a collision during the past 50-150 years of two Kuiper Belt-like objects of radii $\sim50$\,km.

\end{abstract}
    
    \date{Submitted to ApJ on Aug, 9th 2012}
        
    \maketitle
    
\section{Introduction}
Direct imaging is the appropriate technique for the study of exoplanets with semi-major axis larger than a few astronomical units~\citep{marois08,marois10,kalas08,lagrange09}. As the planetary atmospheric thermal emission or scattered light is detected, detailed multi-band photometry or spectrometry can be used to probe the atmospheric composition and physical properties. However, these studies are challenging given the high contrast and small angular separation between a star and planet. In some systems the presence of a planet before it is detected directly can be suggested by the geometry of a circumstellar debris disk. For example, \citet{wyatt99}, \citet{kalas05} and \cite{quillen06} had predicted the likely existence of a planet around Fomalhaut and \citet{mouillet97} of a planet around $\beta$-Pictoris.

In the case of Fomalhaut~\citep[$440\pm40$\,Myr, 7.7\,pc,][]{mamajek12,Leeuwen07}, a candidate planet was announced by \citet[hereafter K08]{kalas08}. Surprisingly, the candidate was not detected in deep near infrared images in~H and~Lp bands, but rather in Hubble Space Telescope (HST) images in the visible where planets are not expected to emit much thermal light. The K08 planet model that best fit the~2008 photometry is a~$<3M_\mathrm{J}$ Jovian planet surrounded by a large circumplanetary disk; the observed optical light is mostly scattered by the disk, not by the planet itself.  In this model, H$\alpha$ emission~(dust accretion or hot planetary chromosphere) explains the unusual~50\% variability of the reported flux at~$0.6\,\mu$m over a two year time interval. Based on their astrometric mesurements, \citet{kalas10} also announced that Fomalhaut~b is likely to cross the dust ring.

A few years later, as they did not detect the object at~4.5$\,\mu$m with Spitzer, \citet{janson12} concluded that "there is almost certainly no direct flux from a planet contributing to the visible-light signature" and they proposed an optically thin dust cloud with or without a central object in the super-Earth regime to explain the~K08 photometry. \citet{kennedy11} also rejected direct detection of massive planets and explain the photometry at 0.6-0.8$\,\mu$m to be a consequence of a swarm of satellites around a 2-100\,$M_\mathrm{Earth}$ planet. 

Motivated by the controversial status of~Fomalhaut~b within the community, including even doubts of its actual existence, we decided to conduct an independent analysis of the~HST public data that were recorded in 2004, 2006, 2009 and 2010. After describing the observational method in~\S\,\ref{sec : obs} and our data reduction in~\S\,\ref{sec : reduc}, we analyze the images to confirm that Fomalhaut~b is a convincingly real detection and that it is gravitationally bound to the star. In~\S\,\ref{subsec : astro} we study various possible orbits to determine if the current astrometry can confirm or reject a dust belt crossing trajectory~\citep[such as one announced by][]{kalas10}.  In~\S\,\ref{sec : res} we estimate the object's photometry and possible origin as a circumplanetary disk around a planet or the aftermath of a collision of two Kuiper Belt-like objects, while considering that Fomalhaut~b is not~(\S\,\ref{subsec : psf}) or is~(\S\,\ref{subsec : ext}) spatially resolved in the HST images.

\section{Observations}
\label{sec : obs}
The data that we consider in this paper were obtained with HST with the Advanced Camera for Surveys~(ACS) in~2004 and~2006~(programs 10390 and~10598), with the wide-field-camera 3 (WFC3) in 2009 (program 11818), and the Space Telescope Imaging Spectrograph instrument~(STIS) in~2010~(program~11818).
\begin{table*}[!ht]
\begin{center}
\begin{tabular}{lccccccc}
\tableline
\tableline
Date (UT)&Instrument&Spot diam.&Filter&Im.&Exp.&Roll&FOV rot.\\
 & &  (arcsec) &  & &  (sec) &  & (deg)\\
\tableline
2004 Oct. 25-26&ACS/HRC&1.8&F606W&112&5615&3&8.0\\
2006 Jul. 14&ACS/HRC&1.8&F435W&9&6525&3&5.8\\
2006 Jul. 15-16&ACS/HRC&3.0&F435W&9&6435&3&5.8\\
2006 Jul. 19-20&ACS/HRC&3.0&F606W&28&7240&4&6.0\\
2006 Jul. 18&ACS/HRC&1.8&F814W&20&5280&3&6.0\\
2006 Jul. 19&ACS/HRC&3.0&F814W&27&4942&3&6.0\\
2009 Nov. 16 & WFC3/IR&-&F110W&4&4772&4&15.0\\
2010 Jun. 14&STIS/50CORON&2.5&CLEAR&3&630&\multirow{2}[0]{*}{7}&\multirow{2}[0]{*}{157.0}\\
2010 Sep. 13&STIS/50CORON&2.5&CLEAR&16&3000&&\\
\end{tabular}
\caption{\sl  Fomalhaut observing log. Column "Im." gives the number of useful images. Column "Exp." is the total integrated time of the useful images. "Roll" is the number of roll angles in the sequence and "FOV rot." gives the total FOV rotation during the sequence.}   \label{tab : obs}
\end{center}
\end{table*}
The~ACS data were acquired with the High Resolution Channel~(HRC) in its coronagraphic mode with~1.8$^{\prime\prime}$ and~3.0$^{\prime\prime}$ focal plane occulting masks and the F435W ($430\pm50$\,nm), F606W ($595\pm115$\,nm), and~F814W ($825\pm115\,$nm) filters. The F110W (${\bf 1150\pm250}$\,nm) filter was used for the acquisition of the WFC3 data. For the~STIS data, the {\sc 50coron} configuration was used with its clear aperture ($600\pm220\,$nm). For all sequences, images at several roll angles were taken so that the stellar diffraction pattern can be subtracted while keeping the flux of any point sources. Tab.~\ref{tab : obs} gives the dates of the observations, the instrument configurations~(filter, coronagraph), the number of useful images with the corresponding integration time, and the number of roll angles as well as the total rotation of the field of view.

\section{Data reduction}
    \label{sec : reduc}
\subsection{ACS}
We start from the {\tt drz} drizzled images produced by the~ACS pipeline~(geometric distortion, photometry, and cosmic ray calibrations). For each image, we create a map of pixels that deviate by more than 3.5\,$\sigma$ in a 20$\times$20\,pixel box and we replace them with the median value in the box. We multiply each image by the {\sc photflam} of its header to convert the pixel counts to erg s$^{-1}$ cm$^{-2}$ \AA$^{-1}$ arcsec$^{-2}$. For the Fomalhaut~PSF registration, we first start with the 2006 sequence at F606W that is recorded with the 3.0$^{\prime\prime}$ focal plane mask. We align every image maximizing its correlation with the first image of the sequence (Table 1). The correlation is maximized in the annulus  with inner and outer radii of 140 and 200\,pixels where the central vertical band of width 240\,pixel is removed (saturated detector) and where the coronagraphic focal plane bar is masked. We call this optimization area~A. Once the images are aligned to within~0.1\,pixel, the absolute center of the star~PSF is found by median-combining the aligned images and by registering the resulting image to the image center by maximizing in~A the cross-correlation of the diffraction spikes with themselves in a 180$\degree$ rotation of the image about its center. This procedure defines the absolute center within 0.5\,pixel and we call R606 the registered median-image. We then register all images of F606W sequences maximizing their cross-correlation with R606 in~A. For the F435W and F814W sequences, we scale R606 to the corresponding wavelengths~(814nm for F814W and 480nm\footnote{Better match of the diffraction spikes than for 435nm.} for F435W) and call R435 and R814 the resulting images. We then register all F435W and F814W images maximizing the cross-correlation in~A with R435 and R814 respectively. For every sequence listed in~Tab.\,\ref{tab : obs}, we then subtract the stellar speckles.
  
\begin{figure*}[!ht]
\centering
\includegraphics[width=\textwidth]{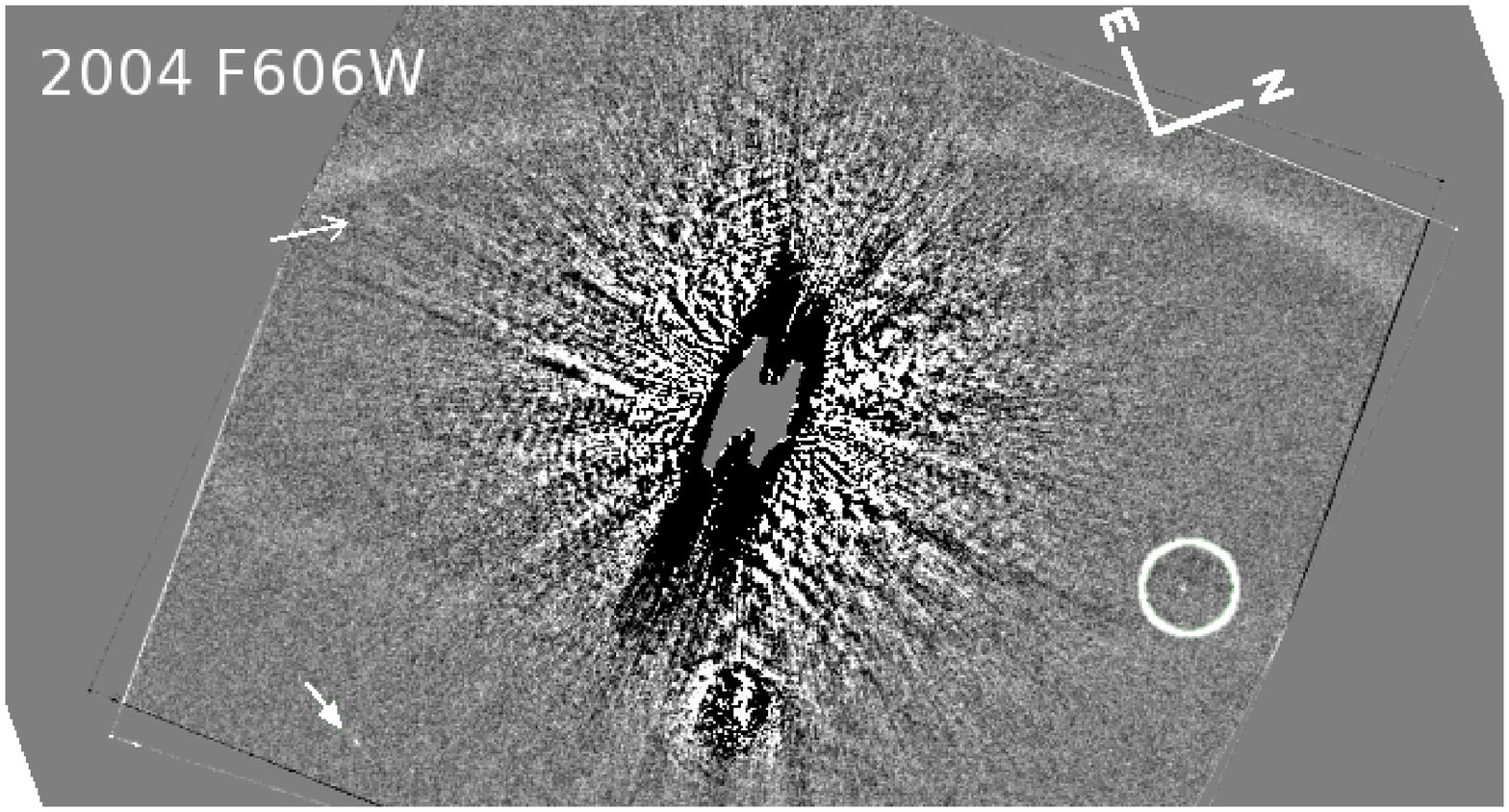}\vspace{.3cm}
\\\includegraphics[width=\textwidth]{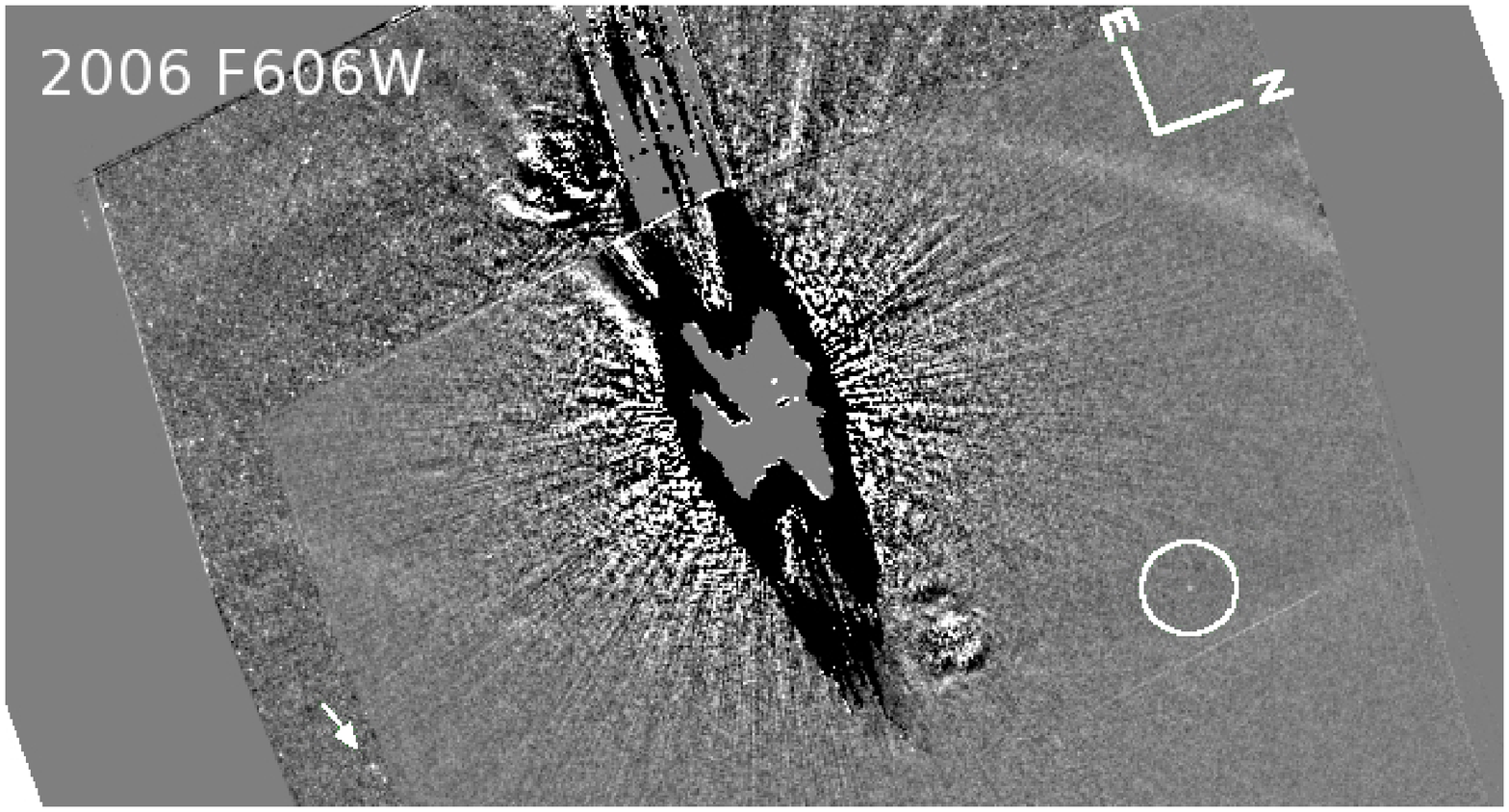}
\caption[]{\sl ACS images of the dust belt and the object~b~(circle) around Fomalhaut at F606W in 2004~(top) and 2006 (bottom). Two arrows point to background sources. The length of the segments giving the East and North orientations is 2.5$^{\prime\prime}$. The intensity scale is linear and it is the same for the two images.}
\label{fig : im}
\end{figure*}

\begin{figure*}[!ht]
\centering
\includegraphics[width=\textwidth]{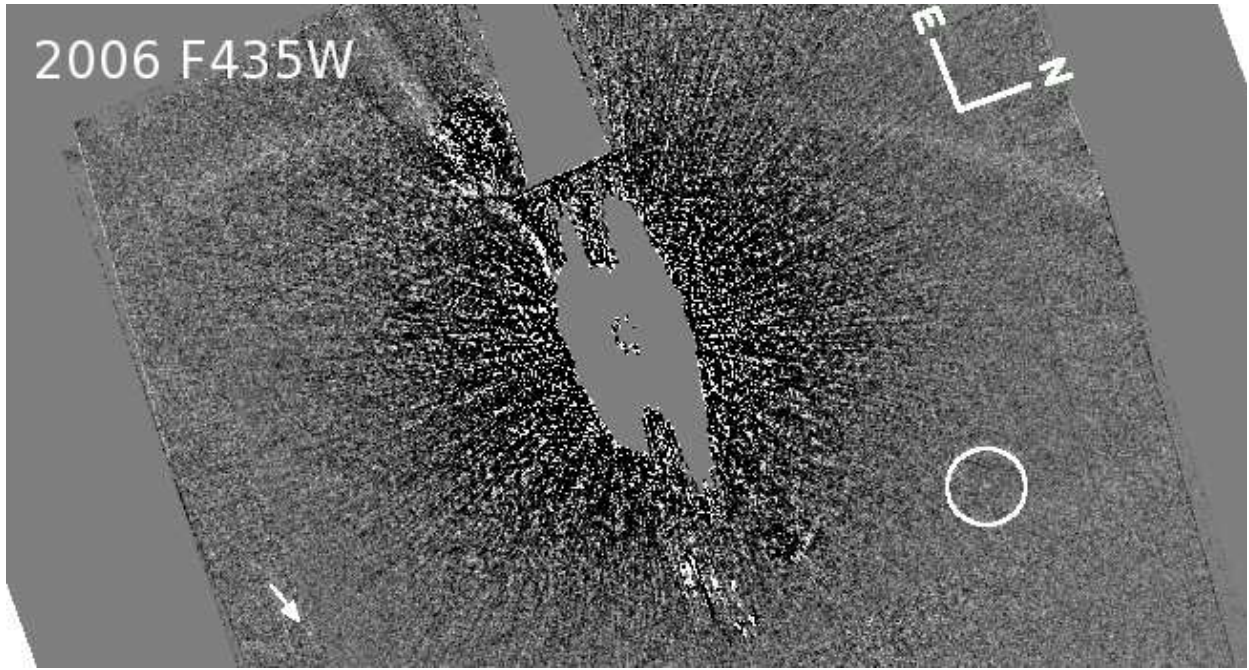}\vspace{.3cm}
\includegraphics[width=\textwidth]{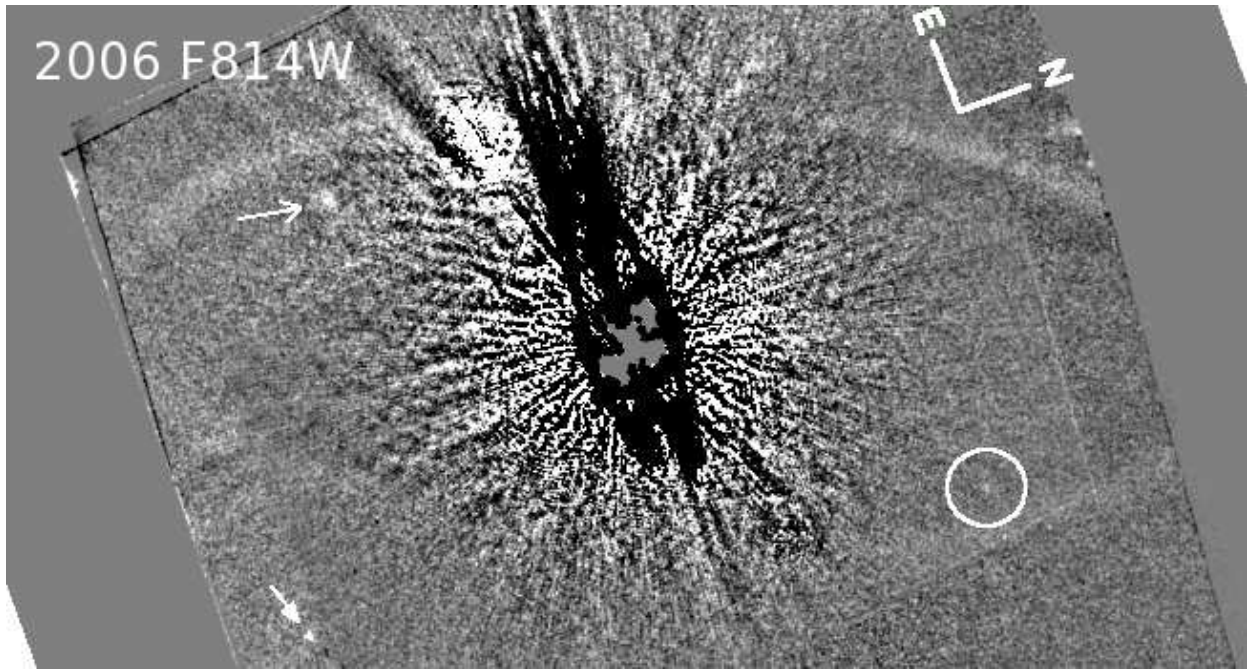}
\caption[]{\sl Same as Fig.\,\ref{fig : im} for F435W (top) and F814W (bottom) ACS images taken in 2006.}
\vspace{1cm}
\label{fig : im1}
\end{figure*}

As the field of view rotates only by a few degrees, there is almost no difference between applying a locally optimized combination of images algorithm~\citep{lafreniere07,marois10b} or a basic angular differential imaging data reduction as described in~\citet{marois06} for all filters. We choose the second procedure which is less time consuming. Considering one of the sequences, we subtract from each image a reference PSF that is the median of all images that were recorded at a different roll angle. We then rotate the images to put North up and median combine them. For the 2006 data, we work out the weighted mean of the reduced images taken with the 1.8$^{\prime\prime}$ and 3.0$^{\prime\prime}$ masks in the same filter. As the fields-of-view do not exactly overlap, the contrast is not the same in all parts of the images~(Figs.\,\ref{fig : im} and \ref{fig : im1}).

\begin{figure*}[!ht]
\centering
\includegraphics[width=\textwidth]{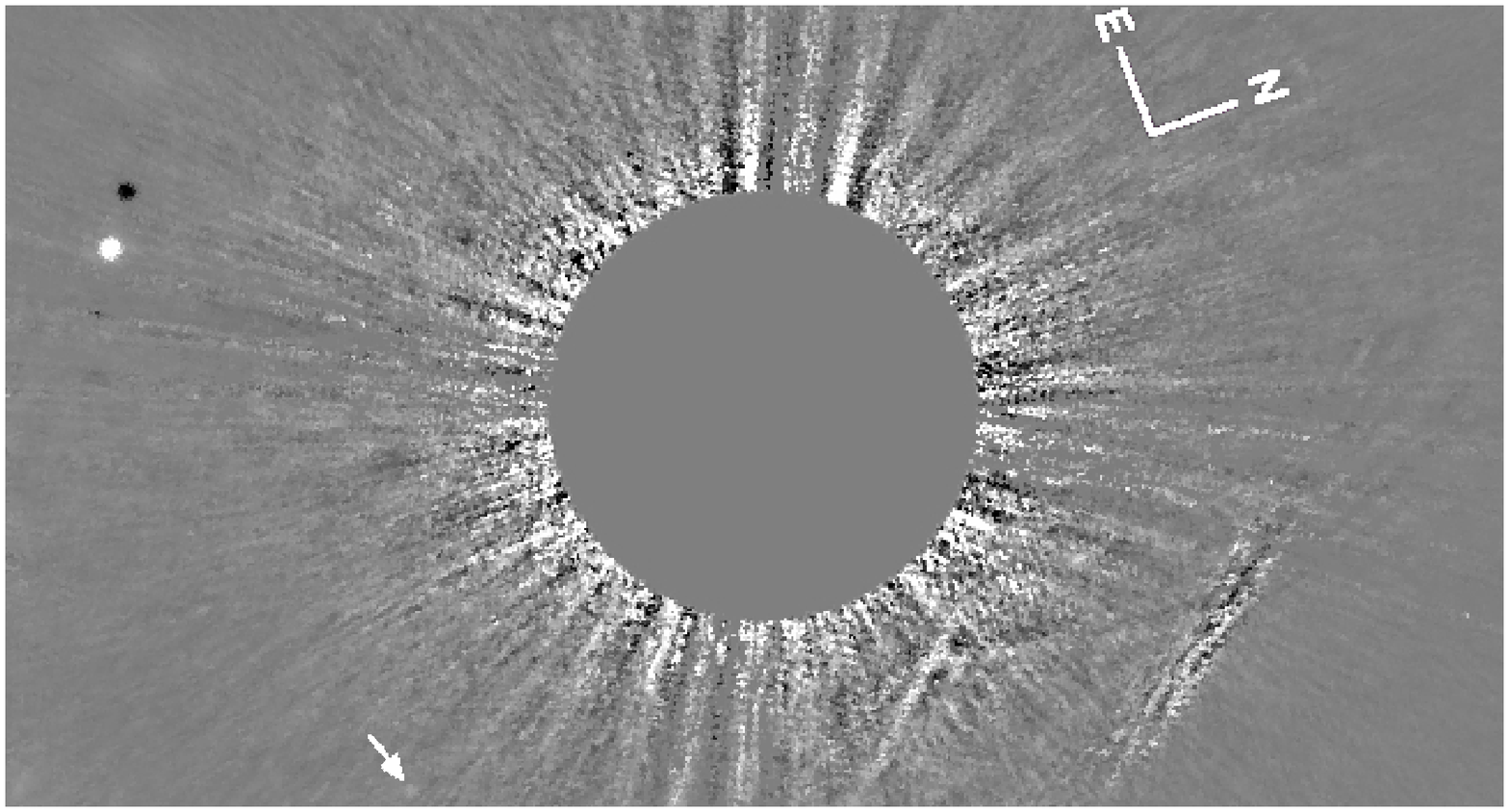}
\includegraphics[width=\textwidth]{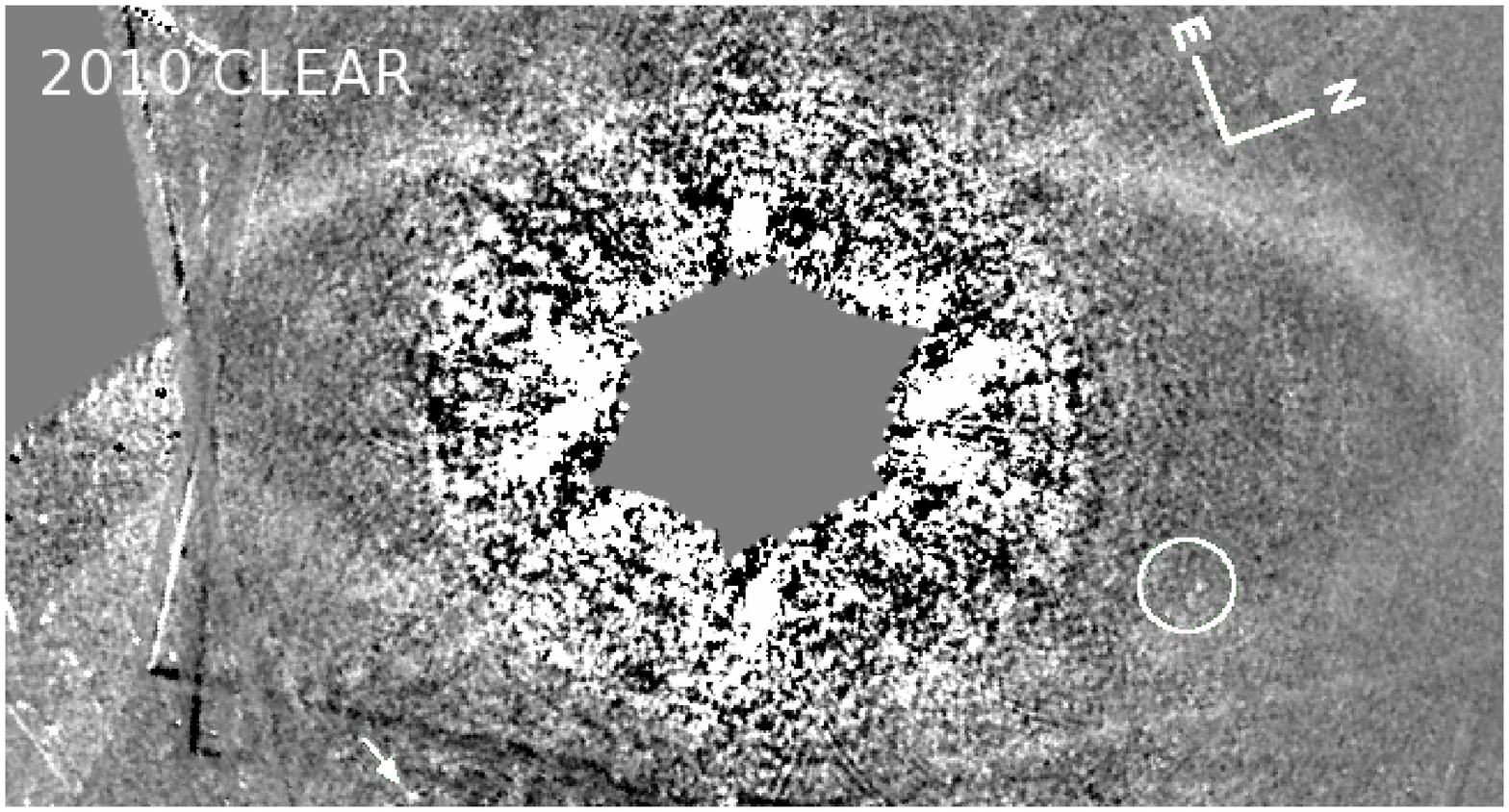}
\caption[]{\sl Same as Fig.\,\ref{fig : im} for the LOCI-processed WFC3 F110W image taken in 2009 (top) and 2010 STIS CLEAR image (bottom).}
\label{fig : im2}
\end{figure*}

\begin{figure}[!ht]
\centering
\includegraphics[width=.48\textwidth]{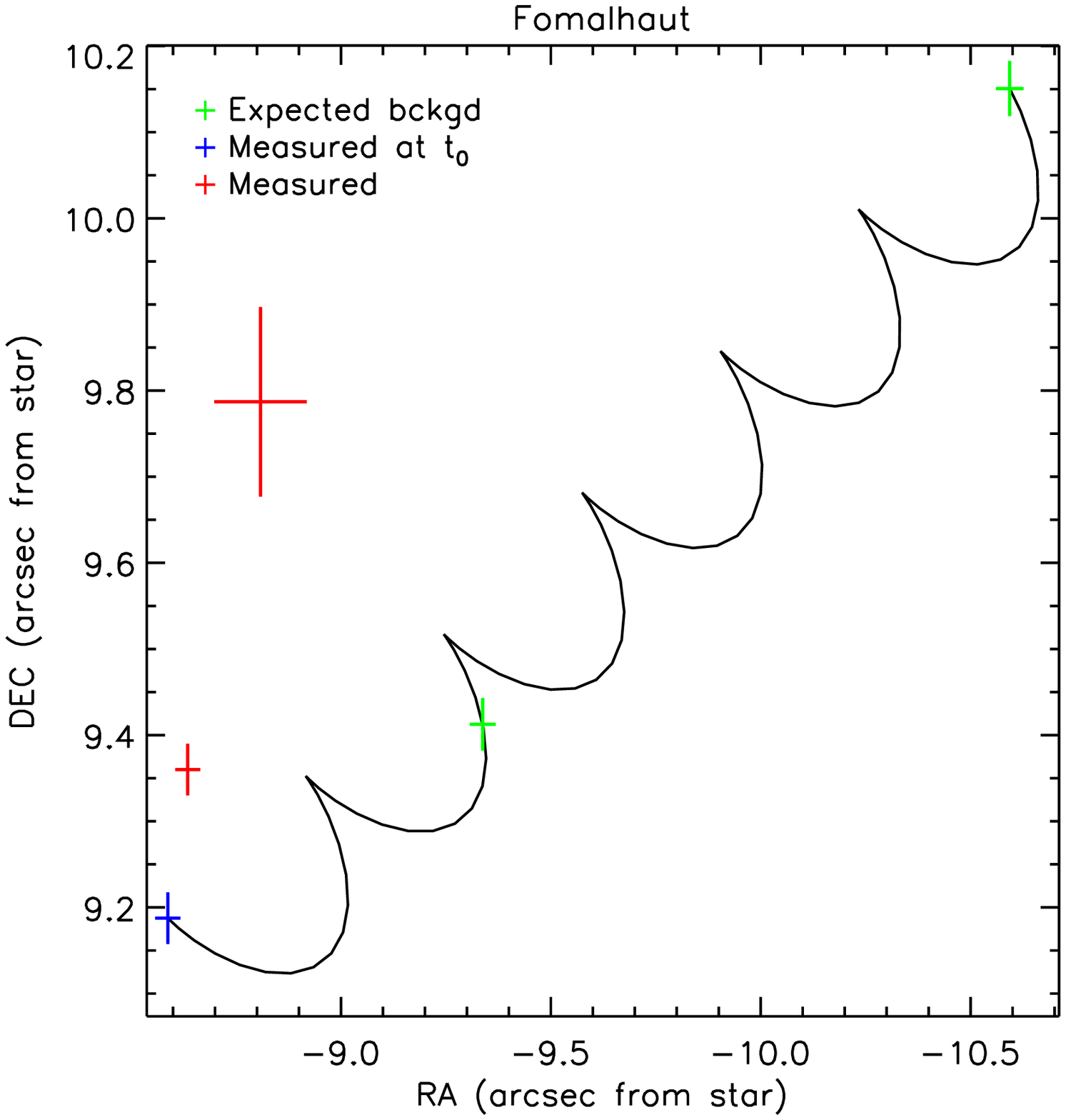}
\caption[]{\sl Fomalhaut~b measured positions in 2004, 2006 and 2010 images~(blue and red crosses) and expected positions for a background source~(green crosses).}
\label{fig : astrograph}
\end{figure}

\begin{figure}[!ht]
\centering
\includegraphics[width=.48\textwidth]{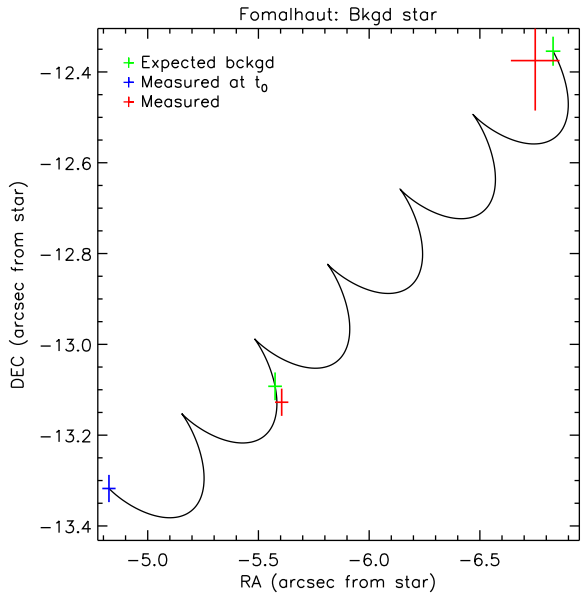}
\caption[]{\sl Measured positions of the South West background source in 2004, 2006 and 2010 images~(blue and red crosses) and expected positions for a background source~(green crosses). Only epochs where Fomalhaut b is detected are shown.}
\label{fig : astrobkgd}
\end{figure}

A point source~(arrow) is detected in all images South West from Fomalhaut~A. An extended object~(red arrow, South East) is also detected in 2004 and in 2006~(F814W). The motion of these two sources are consistent with them being background objects. Fomalhaut~b~(inside the circles) is detected at~F435W, F606W, and F814W with a signal-to-noise ratio of~$\sim3$, 5-6, and~3 respectively and it does not have the same motion as the background sources~(Fig.\,\ref{fig : astrograph}). To confirm that Fomalhaut b is bound and that we detect orbital motion, we have analyzed the astrometry of the South West background source that is at $\sim14$\,arcsec from the star (located at a separation comparable to Fomalhaut b; see~Fig.~\ref{fig : astrobkgd}).  As it fits well the expected positions of a background source~(proper motion and parallax), it means that mis-registration or uncorrected distortions do not bias the astrometry in our images by more than the error bars that we derive.  We thus confirm that Fomalhaut b is a real object orbiting Fomalhaut.

\subsection{STIS}
\label{subsec : stis}
The {\tt sx2} images that are provided by the~STIS pipeline~(geometric distortions, photometry, and cosmic ray calibrations) are used for our analysis. The flux density is converted to erg s$^{-1}$ cm$^{-2}$ \AA$^{-1}$ arcsec$^{-2}$ by multiplying each image by the {\sc photflam} of its header and dividing it by the exposure time. The spider spikes are well detected in these images and we register the first image of the sequence maximizing the cross-correlation of the spikes with themselves in a 180$\degree$ rotation of the image about its center. The maximization was done around the spikes~($\pm$2\,pixels) between~100 and~230\,pixels from the star. We then register the other images maximizing the cross-correlation with the first image in the 200\,pixel-radius disk where the 160\,pixel central vertical band and the 30\,pixel central horizontal band are masked. As the roll angles are well spread in the 0-157$\degree$ interval, we apply a locally optimized combination of images algorithm~\citep{lafreniere07,marois10b} to suppress the stellar diffraction pattern. Using a LOCI algorithm, a~PSF reference image is built for each image of the sequence and it is subtracted. After subtraction, the images are rotated to put North up and they are median-combined. The final image with the detection of Fomalhaut~b is shown in~Fig.\,\ref{fig : im2}.
 
 \subsection{WFC3}
 \label{subsec : wfc3}
 The images that we use in our analysis are the multi-drizzle {\tt drz} F110W images that are provided by the WFC3 pipeline. This pipeline applies geometric distortions, photometry and cosmic ray calibration on all images. Given that the images have been rotated to put North up, the images are first rotated to align the pupil. The first image is registered at the image center using a cross-correlation analysis with a 180 degrees rotated image of itself. The other three images are then registered on the first image using again a correlation analysis. The LOCI algorithm ~\citep{lafreniere07,marois10b} is then applied inside 20 pixels thick annulus without any pixel masking. The subtracted images are then rotated to have North up and are median combined. Due to a bright diffraction artifact, Fomalhaut b is not detected at F110W~(see Fig.\,\ref{fig : im2}).

\section{Data analysis}
\label{sec : res}

\subsection{Belt geometry}
The geometrical properties of the belt have been discussed previously \citep{kalas05,acke12,boley12} and it is beyond the scope of the present paper to refine them. However, we find an eccentric belt that reproduces the images, with an eccentricity  e=0.10-0.11, a radius between 136-148\,AU, a longitude of ascending node  156.5-157$\degree$, an argument of periapsis 35$\degree$, and an inclination of 67$\degree$. All the parameters are in good agreement with the published values of~K08~(31$\pm6\degree$ for the argument of periapsis unlike the 1$\pm6\degree$ found by \citet{acke12}). We did not use a mathematical fit to optimize values of parameters and our best visual fit is only used to estimate the belt geometry in our images.

\subsection{Astrometry}
\label{subsec : astro}
We use the Tiny Tim~\citep{krist11} tool that generates HST template~PSFs to build a model of a point source in our images at the position of Fomalhaut~b to accurately estimate its astrometry and photometry. We consider that Fomalhaut~b is seen in scattered light. Thus, in the Tiny Tim tool, we choose a source which the spectrum is a blackbody with temperature~8751\,K,~\citep[][]{difolco04}. We simulate the images prior to the speckle suppression registering them at the positions where they were recorded on the detector to account for the ADI/LOCI effects, the rotation we apply to put north up, and the weights of the weighted means for the 2006 data. We then adjust the position and flux of the template to subtract from the image to minimize the residual noise in a~0.25$^{\prime\prime}$-radius aperture centered on Fomalhaut~b for the ACS and STIS data. Although the template is close to the real image, the Tiny Tim tool cannot include all variations of the~PSF over the detector.  That is why we choose the~0.25$^{\prime\prime}$ aperture (10\,ACS pixels) as it is large enough to minimize the impact of these approximations; and it is small enough to minimize the impact of the surrounding noise. The positions we derive from the fit are given in Tab.\,\ref{tab : astro}. Note that the uncorrected geometrical distortions induce a~0.01 and~0.1\,pixel error in~ACS/HRC images~(section 10.3 in the handbook) and STIS images~(section~16.1 in the data handbook) respectively. The uncorrected distortions are thus negligible with respect to the fitting errors.
\begin{table*}[!ht]
\begin{center}
\begin{tabular}{lccccc}
\tableline
\tableline
\multirow{2}[0]{*}{Date}&\multirow{2}[0]{*}{Filter}&\multicolumn{2}{c}{Offset from A to b}&Separation&PA\\
&&{\sc ra} ($^{\prime\prime}$)&{\sc dec} ($^{\prime\prime}$) &($^{\prime\prime}$)&(deg)\\
\tableline
2004 Oct. 25-26&F606W&-8.59$\pm$0.02&9.19$\pm$0.02&12.58$\pm$0.03&316.9$\pm$0.1\\
2006 Jul. 14-15-16&F435W&-8.61$\pm$0.03&9.36$\pm$0.03&12.72$\pm$0.04&317.4$\pm$0.2\\
2006 Jul. 19-20&F606W&-8.64$\pm$0.02&9.36$\pm$0.02&12.73$\pm$0.03&317.3$\pm$0.1\\
2006 Jul. 18-19&F814W&-8.64$\pm$0.03&9.36$\pm$0.03&12.73$\pm$0.04&317.3$\pm$0.2\\
2010 Jun. 14-Sep. 13&CLEAR&-8.81$\pm$0.07&9.79$\pm$0.07&13.17$\pm$0.10&318.0$\pm$0.4\\
\end{tabular}
\caption{\sl Fomalhaut~b astrometry with respect to Fomalhaut~A.}   \label{tab : astro}
\end{center}
\end{table*}
K08 measured in their images that Fomalhaut~b is at $[${\sc ra}, {\sc dec}$]=[$-8.62$^{\prime\prime}$, 9.20$^{\prime\prime}]$ and $[$-8.60$^{\prime\prime}$, 9.38$^{\prime\prime}]$ from Fomalhaut~A in 2004 and 2006 respectively. We estimate the difference~$\delta_i$ between K08 positions and ours at epoch~$i$ as
\begin{equation}
\delta_i=\sqrt{\frac{\delta_i \mathrm{{\sc ra}}^2+\delta_i\mathrm{{\sc dec}}^2}{\sigma_{\mathrm{{\sc ra}, i}}^2+\sigma_{\mathrm{{\sc dec}, i}}^2}}
\label{eq : diffpos}
\end{equation}
 where $\delta_i\mathrm{{\sc ra}}$ and $\delta_i\mathrm{{\sc dec}}$ are the difference between K08 measurements and ours of the offset along the West-East direction and the South-North direction respectively.~$\sigma_{\mathrm{{\sc ra}, i}}$ and $\sigma_{\mathrm{{\sc dec}, i}}$ are our error bars~(K08 give no error bars). We find that our positions are within~1.5\,$\sigma$ of~K08 positions at the two epochs 2004 and 2006~(i.e. $\delta_i\lesssim1.5$). The difference with~K08 could result from a different registering technique or from differences in the~ACS pipeline that have been upgraded since~2008.
 
 As we have three epochs close in time and large error bars for the 2010 data, we cannot strongly constrain the orbital parameters. We then consider only two Keplerian orbits -- one that crosses the dust ring and a second that does not -- and compare expected and measured positions.
\begin{figure*}[!ht]
\centering
\includegraphics[width=.6\textwidth]{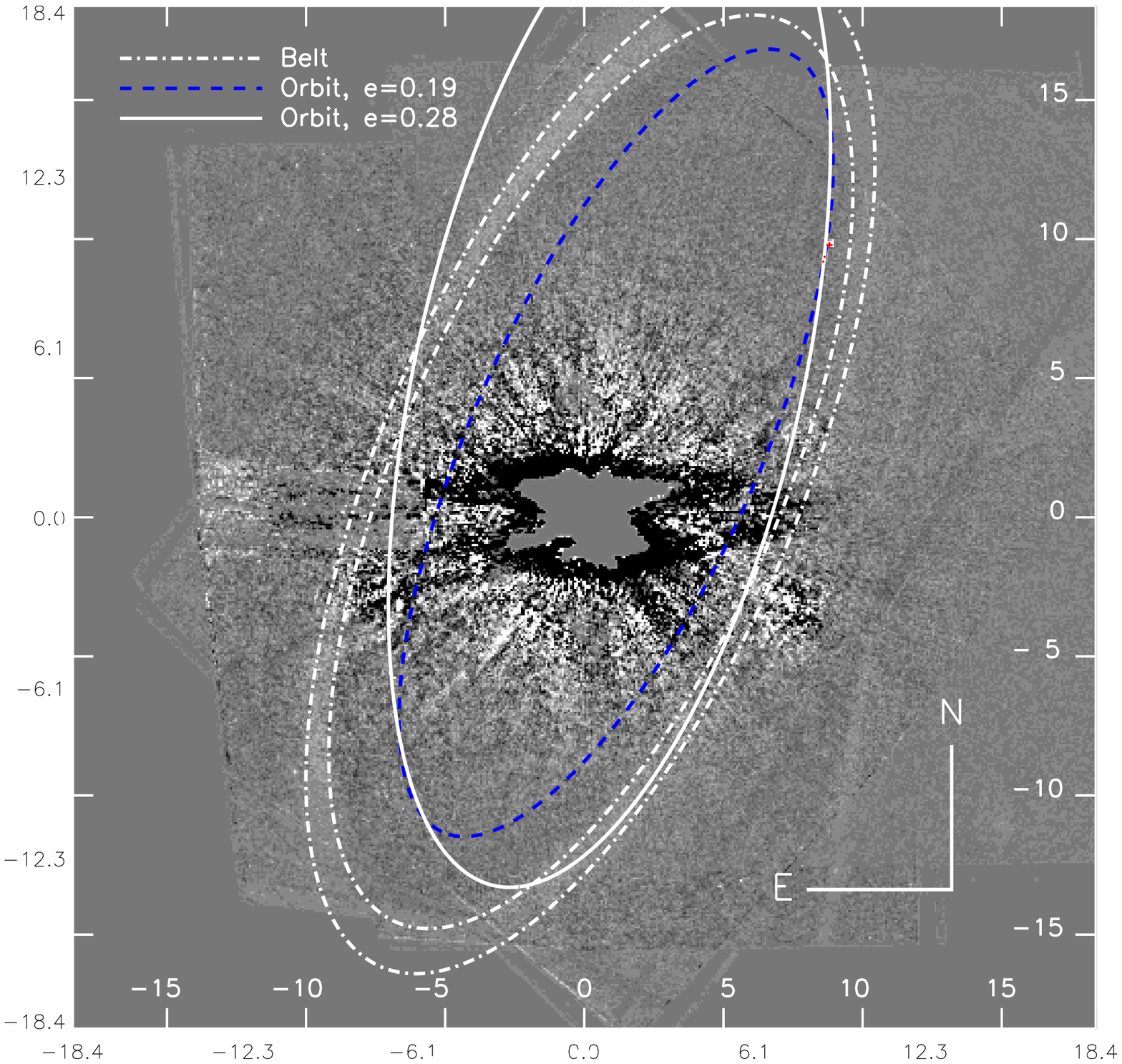}\\
\includegraphics[width=.6\textwidth]{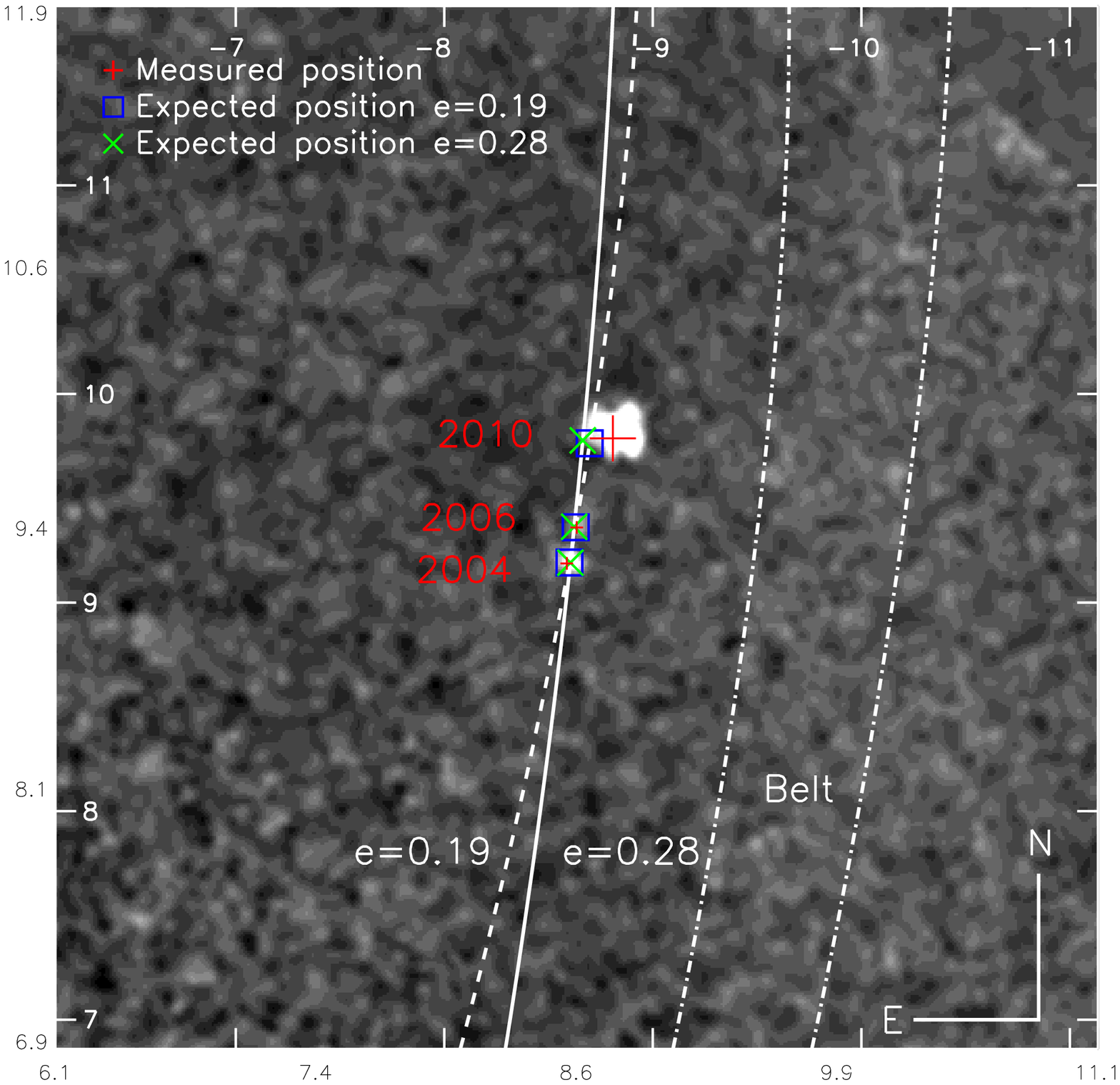}
\caption[]{\sl Two trajectories with eccentricity 0.19~(dashed lines) or 0.28~(full line) that fit the Fomalhaut b positions within $1.5\,\sigma$.}
\label{fig : traj}
\end{figure*}

The first orbit is a~0.19 eccentric orbit with a~118\,AU semi-major axis, a~156$\degree$ longitude of the ascending node, a~70$\degree$ inclination, and a~2$\degree$ argument of periapsis. The orbit does not cross the dust ring and is represented in dashed lines in Fig.\,\ref{fig : traj} where the dust belt is bound by dashed-dotted lines. We use~Eq.\,\ref{eq : diffpos} replacing K08 positions with the expected positions of Fomalhaut~b on the keplerian orbit to estimate the differences $\delta_i$ between the expected positions and our measurements at each epoch~$i$~(2004, 2006, and 2010). Then, we estimate the total difference as $\delta=\sum_i\delta_i$.  We find that the expected positions are $1.1\,\sigma$ from the measured positions~($\delta=1.1$). The second keplerian orbit~(full lines) we consider has an eccentricity 0.28, a semi-major axis 145\,AU, a longitude of the ascending node~167$\degree$, an inclination~67.5$\degree$, and an argument of periapsis~8$\degree$. The difference between expected and measured positions is~1.5\,$\sigma$. If Fomalhaut~b follows this orbit, it was inside the dust belt $140$ years ago at~$\sim1$\,AU from the center of the belt which the full vertical height is~$h_r\sim3.5$\,AU~\citep{kalas05}.  Other trajectories at less than~$2.4\,\sigma$ from the observations put Fomalhaut~b inside the belt $\sim50$ years ago. Some of these trajectories are highly eccentric and may be consistent with results proposed by~\citet{kalas13} and \citet{graham13} although we were not able to find all the parameters of their best fit. Thus, new data are required to conclude whether Fomalhaut~b trajectory does or does not cross the belt.

\subsection{Fomalhaut b as a point source}
\label{subsec : psf}
We consider Fomalhaut~b as a point source in this section. We estimate its photometry and compare our results with~K08 fluxes~(\S\,\ref{subsubsec : photpsf}). We then examine models discussed by~K08 and~J12~(\S\,\ref{subsubsec : cloudpsf} and~\S\,\ref{subsubsec : diskpla}).

\subsubsection{Photometry}
\label{subsubsec : photpsf}
For each filter and epoch, we derive the photometry by integrating the flux density of the~PSF template that best fits the data~(\S\,\ref{subsec : astro}). As the PSF template is generated in a 2.5$\times$2.5\,arcsec$^2$ image, we use a 1.25$^{\prime\prime}$ radius aperture for the ACS data and a 1$^{\prime\prime}$ radius for the STIS data. The fractions of the PSF integrated energy inside these apertures are 0.960, 0.961, 0.918, and 0.996 for the F435W, F606W, F814W \cite{sirianni05}, and CLEAR/STIS (STIS handbook, chap.~14/CCDClearImaging) filters respectively. The F110W flux upper limit is derived by estimating the 5\,$\sigma$ noise in the area where Fomalhaut b is expected to be located, after convolving the image by a 0.4~arcsec diameter aperture (aperture matching the WFC3 {\sc photplam} parameter). To convert the estimated flux densities~$F_\lambda$ in erg s$^{-1}$ cm$^{-2}$ \AA$^{-1}$ to flux densities $F_\nu$ in erg s$^{-1}$ cm$^{-2}$ Hz$^{-1}$~(i.e. $10^{23}$\,Jy), we use the {\sc photplam} keyword recorded by the ACS, WFC3, and STIS pipelines in the fits headers:
\begin{equation}
F_\nu = F_\lambda\ \mathrm{\sc photplam}^2\ 10^{-18.4768}
\end{equation}
The resulting flux densities ($\mu$Jy) are given in~Tab.\,\ref{tab : flux}.
\begin{table}[!ht]
\begin{center}
\begin{tabular}{lccc}
\tableline
\tableline
Date&Filter&\multicolumn{2}{c}{Flux density ($\mu$Jy)}\\
&&&Kalas\\
\tableline
2004 &F606W&0.63$\pm$0.10&0.61$\pm$0.05\\
2006 &F435W&0.36$\pm$0.09&$<$0.87 (5\,$\sigma$)\\
2006 &F606W&0.43$\pm$0.06&0.29$\pm$0.03\\
2006 &F814W&0.36$\pm$0.07&0.37$\pm$0.04\\
2009 & F110W& $<$1.6 (5\,$\sigma$)& -\\
2010 &CLEAR&0.61$\pm$0.21&-\\
\end{tabular}
\caption{\sl Photometry if Fomalhaut b is a point source.}   \label{tab : flux}
\end{center}
\end{table}
The error bars~$\sigma_\nu$ in percentage are the inverse of the signal-to-noise ratios. In these ratios, the signal is the integrated flux density inside a~0.25$^{\prime\prime}$ radius aperture centered on Fomalhaut~b and the noise is the square root of the total variance of the residual noise after subtraction of the best~PSF in the same area. K08 express their Fomalhaut~b photometry and upper limits in Vega magnitudes. We convert their measurements to~$\mu$Jy~(last column in~Tab.\,\ref{tab : flux}) using the~ACS handbook~(section 5.1.1).

Most of the flux densities~(Tab.\,\ref{tab : flux}) are consistent with~K08 values except our F606W/2006 point which is $\sim2\,\sigma$ brighter~($\sigma$ is the quadratic sum of~K08 error bars and ours). Moreover, our error bars are larger than~K08's ones. Thus, even if we still detect a variability in the F606W filter between 2004 and 2006, it is not as significant~($1.7\,\sigma_G$) as it is in~K08~(5-6$\,\sigma_K$) -- where $\sigma_G$ and $\sigma_K$ are our error bars and~K08 ones respectively. We also find that the flux density measured in the CLEAR filter~(its bandpass roughly corresponds to F435W+F606W+F814W) is consistent with the three~ACS flux densities given the large error bar. Finally, we (marginally) detect Fomalhaut~b at F435W unlike~K08 who have an upper limit.

\begin{figure}[!ht]
\centering
\includegraphics[width=.49\textwidth]{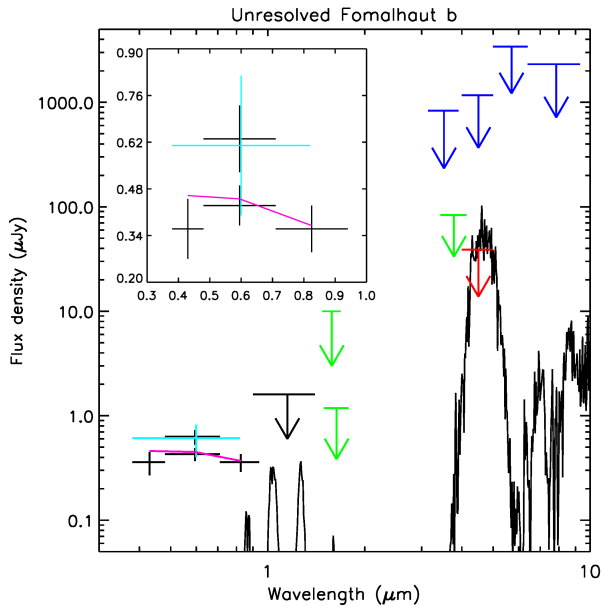}
\caption[]{\sl Fomalhaut~b flux density ($\mu$Jy) for various wavelengths ($\mu$m) in the case the object is unresolved. Crosses correspond to our detections~(four black for ACS and one light blue for STIS). The black arrow is our $5\,\sigma$ upper limit for the flux in the F110W filter. Arrows are $5\,\sigma$ upper limits from the literature: green, red, and blue for K08, J12, and \citet{marengo09} respectively. The solid line represent a cloud-free atmosphere model for a 1\,$M_J$ planet at 400\,Myr. The magenta line give the expected fluxes from a model of a cloud of refractory carbonaceous material (see text).}
\label{fig : sedpsf}
\end{figure}

We plot the photometry of our detections~(crosses) in Fig.\,\ref{fig : sedpsf} along with $5\,\sigma$ upper limits from the literature~\citep[K08, ][and J12]{marengo09} at various wavelengths. 
J12 find that to comply with their~4.5\,$\mu$m upper limit, the planetary mass upper limit is~1$M_J$ at~400\,Myr. Thus, we compare the measurements with a model of a cloud-free atmosphere for a~1\,$M_J$ planet at~400\,Myr with the solar metallicity~\citep{spiegel12}~(full line in Fig.\,\ref{fig : sedpsf}). Our new F110W upper limit is consistent with the expected planet flux at that wavelength given the J12 Spitzer 4.5\,$\mu$m upper limit. It is clear that a planet-only model for Fomalhaut~b is not consistent with the visible observations. K08 proposed two other models: a cloud of dust~(\S\,\ref{subsubsec : cloudpsf}) or a disk of dust around a Jovian-planet~(\S\,\ref{subsubsec : diskpla}). We revisit these two models in light of our updated photometry.

\subsubsection{Cloud of dust}
\label{subsubsec : cloudpsf}
We consider the model introduced in~K08 with a~0.53\,AU diameter cloud composed of dust grains with a differential size distribution $dn/da\propto(a/a_0)^{-3.5}$ where the radius~a goes from $a_\mathrm{min}$ to 1000\,$\mu$m. Using Mie theory, K08 calculate the apparent magnitudes of such a cloud composed of water ice (density=1, $m_\mathrm{ice}$) or refractory carbonaceous material (density=2.2, $m_\mathrm{LG}$) with $a_\mathrm{min}=0.01\,\mu$m~(hereafter $m^{0.01}$) or 8\,$\mu$m~(hereafter $m^8$). The total mass in grains is adjusted such that the integrated light in F814W from the model matches K08's observations~(K08's and our photometry in F814W are in agreement). We convert the Vega magnitudes provided in K08's Tab.\,S3 to flux densities in $\mu$Jy~(Tab.\,\ref{tab : cloudpsf}). The last line gives the error~$\epsilon$ between the expected flux densities~$F_{e,\nu}$ and the observed densities~$F_{\nu}$:
\begin{equation}
\epsilon=\sqrt{\sum_\nu \frac{(F_\nu-F_{e,\nu})^2}{\sigma_\nu^2}}
\end{equation}
\begin{table}[!ht]
\begin{center}
\begin{tabular}{lcccc}
\tableline
\tableline
Filter&$m^{0.01}_\mathrm{ice}$&$m^{0.01}_\mathrm{LG}$&$m^{8}_\mathrm{ice}$&$m^{8}_\mathrm{LG}$\\
\tableline
F435W&0.71&0.58&0.63&0.46\\
F606W&0.55&0.50&0.46&0.45\\
F814W&0.37&0.37&0.37&0.37\\
\tableline
$\epsilon$&2.0&1.3&1.7&1.0\\
\end{tabular}
\caption{\sl Expected flux densities $F_{e,\nu}$ ($\mu$Jy) derived from the cloud models ($m^{0.01}_\mathrm{ice}$, $m^{0.01}_\mathrm{LG}$, $m^{8}_\mathrm{ice}$, and $m^{8}_\mathrm{LG}$) proposed in~K08. The last line gives the difference between~$F_{e,\nu}$ and our measured photometry~$F_\nu$ (see text for details).}
\label{tab : cloudpsf}
\end{center}
\end{table}

K08 reject the possibility that Fomalhaut~b can be explained by one of these cloud models because 1/ they do not detect the object at F435W (they do not reject $m^{8}_\mathrm{LG}$ for this reason),  2/ the red color they observe does not match the model, and 3/ they cannot explain the F606W variability. All these reasons do not apply to our new photometry because 1/ we detect Fomalhaut~b at F435W, 2/ the expected flux densities match the observed flux densities within $1.7\,\sigma$ for three of the four models~($\epsilon<1.7$), and 3/ the F606W variability is not significant in our images. K08 also explain that such a cloud could result from a collision of two planetesimals and that the probability of such an event is lower at the Fomalhaut~b position than closer to the star or closer to the belt. However, as suggested by~J12, the probability of a collision is not the probability of its detection because the speckle noise and the high brightness of the ring may prevent detections of such clouds close to the star and the belt respectively. Moreover, the collision could have occurred inside the ring of dust and the resulting materials could have moved from the ring to the current position of Fomalhaut~b. Finally, K08 argue that such dust clouds would be sheared due to differential gravitational forces and rapidly spatially resolved by~HST. However, assuming a cloud with diameter~0.5\,AU~(maximum size for an unresolved source)  only subject to gravitational forces from the star and no initial velocity, we find that its image would be larger than~2\,pixels~($\sim${\sc fwhm}) and~4\,pixels after $\sim$100\,years and $\sim$200\,years respectively. It would take $\sim$500\,years to shear the cloud of dust so that it could be spatially resolved in the HST images with no doubt. Thus, we find no strong arguments to reject K08 models of a dust cloud with radius $\sim0.5\,$AU, composed of water ice or refractory carbonaceous small grains, and younger than~$\sim500\,$years.  We over plot a line that gives the expected fluxes for the $m^{8}_\mathrm{LG}$ model in Fig.~\ref{fig : sedpsf}.

\subsubsection{Material surrounding a Jupiter-like planet}
\label{subsubsec : diskpla}
A second scenario proposed by~K08 is an unseen Jovian planet surrounded by a disk of dust with a radius of~16-35 planet radius. As the K08 photometry is close to ours and K08 only work out rough numbers (they could not constrain all the parameters with only two photometric points), the 16-35 planet radius disk surrounding an undetected Jupiter-like planet is consistent with our photometry. J12 reject this model because 1/ it does not explain the F606W variability and 2/ the belt geometry would be strongly affected considering a ring-crossing orbit for Fomalhaut~b~\citep{kalas10}. As we do not find a significant F606W variability and our new astrometry cannot reject an orbit that does not cross the ring, we cannot rule out this model using the~K08 arguments. J12 also consider that if the spin of the star is aligned with the plane of the disk, the north-west side of the disk is closer to Earth than the south-east side. In that case, Fomalhaut~b is between its star and Earth in the radial direction and J12 claim that it would be difficult to explain how an optically thick disk can reflect so much light towards Earth. It is true if we observe the non-illuminated side of the disk but we can imagine an inclined disk such that we observe the illuminated part of the disk even if Fomalhaut~b stands between Fomalhaut and Earth.

\subsection{Is Fomalhaut b resolved?}
\label{subsec : ext} 
\subsubsection{Extended source vs PSF}
\label{subsubsec : size}
Given that a possible model for Fomalhaut~b involves a cloud of dust, it would be possible that object has slowly expanded in time. We test here the possibility that the Fomalhaut~b images are slightly spatially resolved.

First, we combine all the Fomalhaut~b ACS images weighting the images by the SNRs of the detections~(linear and quadratic weighting give very similar results), and we fit a 2D-Gaussian function to the combined image. The best Gaussian function~FWHM is~$6\pm 1$\,pixels.

Then, we test how our processing can widen the image of a point-like source. For each filter/epoch of ACS observations, we extract a small subimage close to Fomalhaut~b (at 50 pixels maximum from Fomalhaut~b). We add this noise to the~PSF templates generated in~\S\,\ref{subsec : astro} adjusting the noise level to reach the same~SNRs as we have for the Fomalhaut~b detections. We combine the four epoch/filter images weighting by the~SNRs and we fit a 2D-Gaussian function to the combined image. Applying this analysis for noises picked at eight different locations in each filter/epoch image, we find the PSF~FWHM estimation is~$2.8\pm0.5$\,pixels. We repeat the same full analysis replacing the~PSF templates by the detected South-West background source images, and we find the background source image~FWHM is~$3.8\pm0.5$\,pixels. Assuming this source is not spatially resolved, we conclude that our data processing can widen the image of a point-like source by~$\sim1\pm0.7$\,pixel.

We now model an extended object assuming a uniform intensity distribution over a disk with radius~$R$. We convolve the object model by the~PSF templates~(\S\,\ref{subsec : astro}) and obtain the object image templates for all epochs/filters.  We adjust the~SNRs of the detections adding noise subimages picked around the Fomalhaut~b images. We combine the images accounting for the~SNRs and fit a 2D-Gaussian function. FWHMs found for sources with~$R$ between $0.39$ and $0.78$\,AU are at less than $1\,\sigma$~(estimated from noises picked at eight different locations) from the $6$\,pixel FWHM measured for the Fomalhaut~b image.

Finally, we find that the Fomalhaut~b image FWHM is ~$\sim2\,\sigma$ from the widening induced by our data processing, suggesting that Fomalhaut could be resolved, but it is not yet conclusive. We also find that a basic model of an extended source could explain the measured Fomalhaut~b extension. It is clear this low SNR analysis is not sufficient to fully conclude whether Fomalhaut~b is or is not spatially extended; new observations are required. However, in the rest of section~\S\,\ref{subsec : ext}, we consider an extended source which the intensity distribution is a uniform disk with radius~0.58\,AU~(3\,ACS pixels).

\subsubsection{Unlikely an instrumental effect}
\label{subsubsec : instru}
Assuming the image is resolved, we investigate what instrumental effect or data processing could explain such an extended source.

The ACS/HRC PSF is contaminated by a halo for red sources, especially at F814W (section~5.1.4 in the ACS handbook). The halo which adds to the "normal" PSF has a diameter (42-2.36\,$\lambda$)\,pixels and contains a total fractional intensity $2\,(\lambda-0.45)^3$ for the wavelength~$\lambda$ in microns. A~10\,pixel diameter halo requires a dominant flux at $\lambda\sim11\,\mu$m from this expression, which does not make sense because it is well outside the sensitive bandpass of the detector. Moreover, even if the signal-to-noise ratio is low, we do not observe in the F814W image a PSF plus a halo but only an extended image.

A second explanation for such an image could be a misregistering of the raw images. In that case, after the rotations that put north up in the~ADI process, all the Fomalhaut~b images would not fall at the exact same position, resulting in a blurred image. If this happens, any source in the field of view would be affected the same way. This effect is included in the estimated widening induced by our processing~(\S\,\ref{subsubsec : size}).

The last instrumental effect that we foresee is a differential geometric distortion of $\gtrsim1$\,pixel at the Fomalhaut~b position between the images of a same sequence. The~ACS pipeline corrects for the distortions with an accuracy~0.01\,pixels~(section~10.3 in the ACS handbook). Thus, it would require differential distortions 100 times larger than the pipeline accuracy at the Fomalhaut~b position but almost no distortions at the background source position which is roughly at the same angular separation from the star. This scenario seems very unlikely.

Finally, we find no instrumental effects that could explain the possible spatial resolution of Fomalhaut~b in our images. Since the current paper was submitted, \citet{kalas13} mentioned that Fomalhaut~b image appears slightly extended in the 2012 images, which is qualitatively consistent with our analysis of the three earlier epochs. However, we insist that more observations with higher~SNR are needed to establish whether or not Fomalhaut~b is extended in the HST images.

\subsubsection{Photometry}
\label{subsubsec : photext}
For each filter/epoch, we consider the template~$T_o$ for a~1.16\,AU diameter object. We adjust its flux to minimize the residual noise in a~0.25$^{\prime\prime}$ radius aperture when we subtract it from the observations. We follow the steps described in~\S\,\ref{subsubsec : photpsf} to convert the flux densities to~Jy and estimate the error bars. The results are given in Tab.\,\ref{tab : fluxext}.
\begin{table}[!ht]
\begin{center}
\begin{tabular}{lcc}
\tableline
\tableline
Date&Filter&Flux density ($\mu$Jy)\\
&&\\
\tableline
2004 &F606W&0.91$\pm$0.10\\
2006 &F435W&0.63$\pm$0.09\\
2006 &F606W&0.60$\pm$0.08\\
2006 &F814W&0.48$\pm$0.09\\
2009 & F110W& $<$ 1.6 (5\,$\sigma$)\\
2010 &CLEAR&1.04$\pm$0.20\\
\tableline
]
\end{tabular}
\caption{\sl Photometry if Fomalhaut b is~1.16\,AU large.}
\label{tab : fluxext}
\end{center}
\end{table}

\begin{figure}[!ht]
\centering
\includegraphics[width=.49\textwidth]{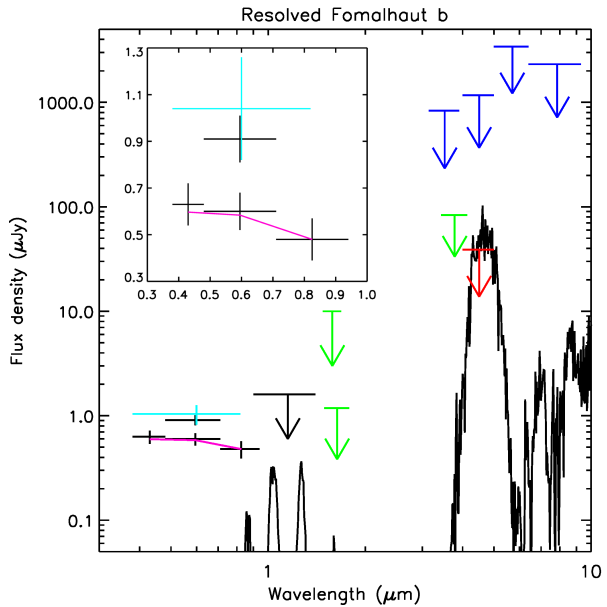}
\caption[]{\sl Same as Fig.\,\ref{fig : sedpsf} in the case the object is resolved. The magenta line give the expected fluxes from a cloud of refractory carbonaceous material (see text).}
\label{fig : sedext}
\end{figure}

As expected, the fluxes are larger than in the case of a point source. Moreover, the flux variation at F606W is larger than in the point source case but it is still less than 2.5\,$\sigma_G$, thus not yet significant. An unfortunately situated speckle at less than 3\,pixels from Fomalhaut~b could explain this variation. Finally, the flux density measured in the large band of STIS is consistent with the average flux density measured in the~ACS filters within~1.8\,$\sigma_G$. In the case we resolve Fomalhaut~b, we plot the photometry of our detections~(crosses) in~Fig.\,\ref{fig : sedext}.
In Fig.~\ref{fig : sedext}, we add the fluxes derived from the $m^{8}_\mathrm{LG}$ K08 model of a cloud of refractory carbonaceous material~(\S\,\ref{subsubsec : cloudpsf}). We multiply the three fluxes by 0.47/0.38, i.e. we adjust the F814W flux and assume the ratios between filters are the same. The model seems to be in good agreement with the data.

In the case of a spatially resolved Fomalhaut~b, we propose one basic model that assume that Fomalhaut~b is the result of the collision of two Kuiper belt objects~(\S\,\ref{subsubsec : kbocol}) and we adapt a model of circumplanetary swarm of satellites~(\S\,\ref{subsubsec : swarm}) proposed by \citet{kennedy11}.

\subsubsection{Collision of Kuiper belt objects}
\label{subsubsec : kbocol}
In this section, we propose a basic model to roughly estimate the size and the amount of light that is scattered by a cloud of dust produced by the collision of two Kuiper belt objects~(KBO). The objective is not to derive the exact radius, mass and velocity of the KBOs that could create Fomalhaut~b but to show that the collision of two KBOs is not completely inconsistent with the observations. First, we estimate the total grain mass that can explain the fluxes received from Fomalhaut~b. Then, we show that the amount of dust can be the result of a collision of two 50\,km radius colliders. We evaluate the rate of collisions of two such KBOs around Fomalhaut. Finally, we estimate when the collision may have occurred to reproduce the size of the Fomalhaut~b images.

If the particles of dust are spheres with radius~$a$ and if the cross section of the particles equals their geometric albedo, the mass~$M_d$ of a cloud of dust that lies at a distance~$D$ from its star is~\citep{jura95}
\begin{equation}
M_d \gtrsim \frac{16\,\pi}{3}\,\rho\,D^2\,a\,\frac{L_{sc}}{L_*}
\label{eq : md}
\end{equation}
where $\rho$ is the mass density of the dust grains, and $L_{sc}$ and $L_*$ are the luminosity of the light scattered by the cloud and the luminosity of the star respectively. Instead of estimating the ratio of the luminosities, we work with the fluxes~$F_{sc}$ and~$F_*$ received at the telescope. We estimate~$F_{sc}$ from the measured fluxes~($F_i$) in the ACS filters~(i=F435W, F606W, and F814W, Tab.\,\ref{tab : fluxext})
\begin{equation}
F_{sc} = \sum_i\,F_i \Delta\nu_i
\label{eq : lsc}
\end{equation}
with $\Delta\nu_i$ the bandwidths of the filters. Our estimation of~$F_{sc}$ only includes the scattered energy in the F435W, F606W, and F814W bandpasses.~$F_*$ has to be calculated for the same bandpass. Assuming a Planck law,~$F_*$ in the bandpass [$\lambda_{min}=435-50\,$nm$,\lambda_{max}=825+115\,$nm] is
\begin{equation}
F_* = F_{*,tot}\,\frac{\int_{u_{min}}^{u_{max}} u^3/(\exp{u}-1)\,\mathrm{d}u}{\int_0^{\infty} u^3/(\exp{u}-1)\,\mathrm{d}u}
\label{eq : lst}
\end{equation}
where $u_{min, max}=h\,c/(k\,T\,\lambda_{max, min})$ with the Planck constant~$h$, the speed of light in vacuum~$c$, the Boltzmann constant~$k$, the stellar effective temperature~$T$~\citep[8751\,K,][]{difolco04}, and the stellar flux $F_{*,tot}$ received at the telescope~\citep[8.914e-6\,erg.cm$^{-2}$, ][]{kalas08}. For $D\sim120$\,AU and dust grains with radius~$a=$10\,$\mu$m and~$\rho=2$\,g.cm$^{-3}$, we find from Eqs.\,\ref{eq : md}, \ref{eq : lsc}, and \ref{eq : lst} that the total grain mass needed to reproduce the photometry of Fomalhaut~b is $M_\mathrm{d}\sim4.10^{19}\,$g.

\citet{jewitt12} estimates the mass~$m_\mathrm{e}$ of particles that are ejected after a collision of two KBOs with a mass~$M_\mathrm{kbo}$, a radius~$r$, a density~$\rho$, and a relative velocity~$U$
\begin{equation}
\frac{m_\mathrm{e}}{M_\mathrm{kbo}}=A\,\left[r\,\sqrt{\frac{8\,\pi\,G\,\rho}{3}}\right]^{-1.5}\,U^{1.5}
\end{equation}
where A equals 0.01 and $G=6.67\,10^{-11}$\,m$^3$ kg$^{-1}$ s$^{-2}$ is the gravitational constant. \citet{jewitt12} assumes the particles have radii in the range $0.1\,\mu\mathrm{m}\lesssim a\lesssim0.1$\,m with a power law distribution in radii with index $\sim3.5$~\citep[see also][]{kadono10}. For typical KBOs in the ring, $U$ is the orbital velocity times $h_r/(2D)$ with $h_r$ the full vertical height of the ring at radius $D$. With $h_r\sim3.5$\,AU~\citep{kalas05} at $D\sim120$\,AU, $U$ is close to $60$\,m s$^{-1}$. Assuming KBOs with radius~$r=50$\,km, the total debris mass~$m_\mathrm{e}$ after the collision is roughly 1$\%$ of the mass~$M_\mathrm{kbo}$ of one of the two colliders with a density $\rho=2$\,g\,cm$^{-3}$. Given the approximations in the models, the expected mass of dust~($1\%\,M_\mathrm{kbo}$) that is ejected after a collision of two 50\,km radius KBOs is consistent with the mass estimated from the photometry of Fomalhaut~b~($4\,\%\,M_\mathrm{kbo}$ for a 50\,km radius KBO).

We assume a maximum post-collison outflow velocity at infinity equal to the escape velocity, $r\,\sqrt{8\,\pi\,G\,\rho/3}$. Considering this upper limit, the diameter~$s$ of the cloud is $2\,r\,t\,\sqrt{8\,\pi\,G\,\rho/3}$ at the date~$t$ after the collision and reaches the observed size~$s=1.16\,$AU (\S\,\ref{subsubsec : size}) after $\sim50$\,years, which is then a lower limit to the time since the collision of the putative Kuiper Belt objects. We can also estimate from the expansion expression that, after $\sim150$\,years, the source would have a diameter $\sim3.5$\,AU and would be $\sim10$ times fainter than the current detections assuming the same amount of reflecting dust. It would not be detected in our images. Thus, if Fomalhaut~b is the product of a collision, the event should have occurred between $\sim50$ and~$150$ years ago to be consistent with our detections. This range is consistent with the possible trajectories that put Fomalhaut~b inside the ring of dust $50-150$ years ago~(\S\,\ref{subsec : astro}). As the size of the possible extended source is very approximative, we keep in mind that these numbers are coarse estimations.

Finally, we evaluate the rate of a collision of two $r=50$\,km KBOs inside the ring of dust. We first find the collision time which reads
\begin{equation}
t_\mathrm{col} = \frac{1}{4\,\pi\,n\,r^2}\,\frac{1}{U}
\label{eq : tcol}
\end{equation}
with $n$ the number of KBOs with radius~$r$ per unit volume. To estimate~$n$ we need the mass $M_\mathrm{disk}$ of the debris disk around Fomalhaut. We assume $M_\mathrm{disk}$=40\,$M_\mathrm{Earth}$ because 1/ estimates of the mass of the Sun's early Kuiper belt is 40\,$M_\mathrm{Earth}$~\citep{schlichting11}, and 2/ estimates of the mass of the Vega debris belt is 10\,$M_\mathrm{Earth}$ in objects with radii $<100$\,km~\citep{muller10} whereas the Vega IR luminosity is 4 times fainter than that of Fomalhaut. Simplifying by setting the radii of all the KBOs to 50\,km does not qualitatively alter the collision frequency estimated below. Under these conditions and considering the belt surrounding Fomalhaut~A has a volume~$2\,\pi\,D\,\Delta D\,h_r$ with $\Delta D\sim0.13\,D$~\citep{kalas05}, the number of~KBOs is
\begin{equation}
n\,2\,\pi\,D\,\Delta D\,h_r=\frac{M_\mathrm{disk}}{M_\mathrm{kbo}}\sim2\times10^8
\label{eq : nbkbo}
\end{equation}
Using $U=\pi\,h\sqrt{M_*/M_\odot\,(1\mathrm{AU}/D)^3}$, and Eqs.\,\ref{eq : tcol} and\,\ref{eq : nbkbo}, we can write
 \begin{equation}
 t_\mathrm{col}=\frac{0.13}{2\,\pi}\left(\frac{D}{r}\right)^2\frac{M_\mathrm{kbo}}{M_\mathrm{disk}}\sqrt{\left(\frac{D}{1\mathrm{AU}}\right)^3\frac{M_\odot}{M_*}}
\end{equation}
Finally, the rate~$\kappa$ of collisions of two KBOs with radii $r$ is the ratio of the number of KBOs~(Eq.\,\ref{eq : nbkbo}) to $t_\mathrm{col}$
\begin{equation}
\kappa=\frac{2\,\pi}{0.13}\left(\frac{r}{D}\right)^2\left(\frac{M_\mathrm{disk}}{M_\mathrm{kbo}}\right)^2\sqrt{\left(\frac{1\mathrm{AU}}{D}\right)^3\frac{M_*}{M_\odot}}
\label{eq : rate}
\end{equation}
Eq.\,\ref{eq : rate} with $M_\mathrm{disk}=40\,M_\mathrm{Earth}$, and $M_*=2\,M_\odot$ indicates that $\sim1$ collision of two $r=50$\,km KBOs occurs every century in the ring around Fomalhaut~A. The rate is low enough to explain that we detect only one event around Fomalhaut as each event would be detectable during $\sim200$\,years in our images. At the same time, it is high enough to make such a $\sim50$ to $150$ year-old event plausible.

In summary, we conclude that it is plausible that Fomalhaut~b is a cloud of dust that was produced $\sim50-150$\,years ago inside the dust belt by the collision of two KBOs with radii $\sim50$\,km.

\subsubsection{Circumplanetary satellite swarm}
\label{subsubsec : swarm}
\citet[KW11]{kennedy11} propose a model of circumplanetary satellite swarms that they apply to Fomalhaut~b. They find that the planet mass can be $\sim$2-100\,$M_\mathrm{Earth}$ surrounded by a swarm that lies at~0.1-0.4 Hill radii. The swarm mass would be of the order of a few lunar masses. But these numbers are derived from K08 photometry of an unresolved source.

Here, we use the same model under the same assumptions ~(body size distribution and maximum/minimum body sizes, dust density, etc) but we consider a swarm of satellites with diameter 1.16\,AU, our photometry~(Tab.\,\ref{tab : fluxext}), and a star with age 440\,Myr instead of 200\,Myr~\citep{mamajek12}. We do not describe the model as it is done in~KW11. We only use the meaningful equations to constrain the planet mass and the swarm mass and size following the steps in~\S\,3.3.1 and~\S\,3.3.2 of~KW11. First, we derive the total cross-sectional area of dust~$\sigma_\mathrm{tot}$ from our photometry: $\sigma_\mathrm{tot}= 6.12\times10^{-4}\,\mathrm{AU}^2$, assuming a geometric albedo~0.08, a phase function~0.32~(Lambert sphere at maximum extension from its host star), the star effective temperature~8,751\,K \citep{difolco04}, and a stellar luminosity $6.34.10^{27}$\,W~(K08).

As we assume that we resolve~Fomalhaut~b, we can write $2\,\eta\,R_\mathrm{Hill}=s$, where~$\eta$ is the semimajor axis of the satellites of the swarm relative to the Hill radius~$R_\mathrm{Hill}$ at the Fomalhaut~b separation~(118\,AU) and~$s$ is the swarm diameter. As explained in~\S\,\ref{subsubsec : size}, the size of the extended source s=1.16\,AU is approximate and at F814W, the image of Fomalhaut~b could be reproduced by a source with radius up to s=2.32\,AU. Thus, we consider 1.16\,AU$<s<$2.32\,AU. Using the Hill radius expression~Eq.\,1 in~KW11) and~2 solar masses for Fomalhaut~(KW11), we derive two constraints: $0.61/M_\mathrm{pl}^{1/3}<\eta<1.22/M_\mathrm{pl}^{1/3}$, where $M_\mathrm{pl}$ is the planet mass expressed in Earth masses.

Considering a collision-limited satellite swarm around Fomalhaut~b (i.e. swarm has just started to suffer collisions) that reproduces the observed~$\sigma_\mathrm{tot}$, it imposes a minimum limit for the satellite semimajor axis $\eta>0.29/M_\mathrm{pl}^{0.12}$ for a~440\,Myr system~(i.e. a 440\,Myr collision time, see~KW11 for details).

KW11 also study the collision velocities that are required to destroy a large object at the Fomalhaut~b position. Assuming a steady-state collisional cascade and a two-phase size distribution for the particles, KW11 link the collision velocity to the swarm size~$\eta$ and the planet mass~$M_\mathrm{pl}$. Using their equations in the case of a resolved object we set a constraint that reads $\eta>0.69/M_\mathrm{pl}^{0.46}$~(KW11).

\begin{figure*}[!ht]
\centering
\includegraphics[width=0.8\textwidth]{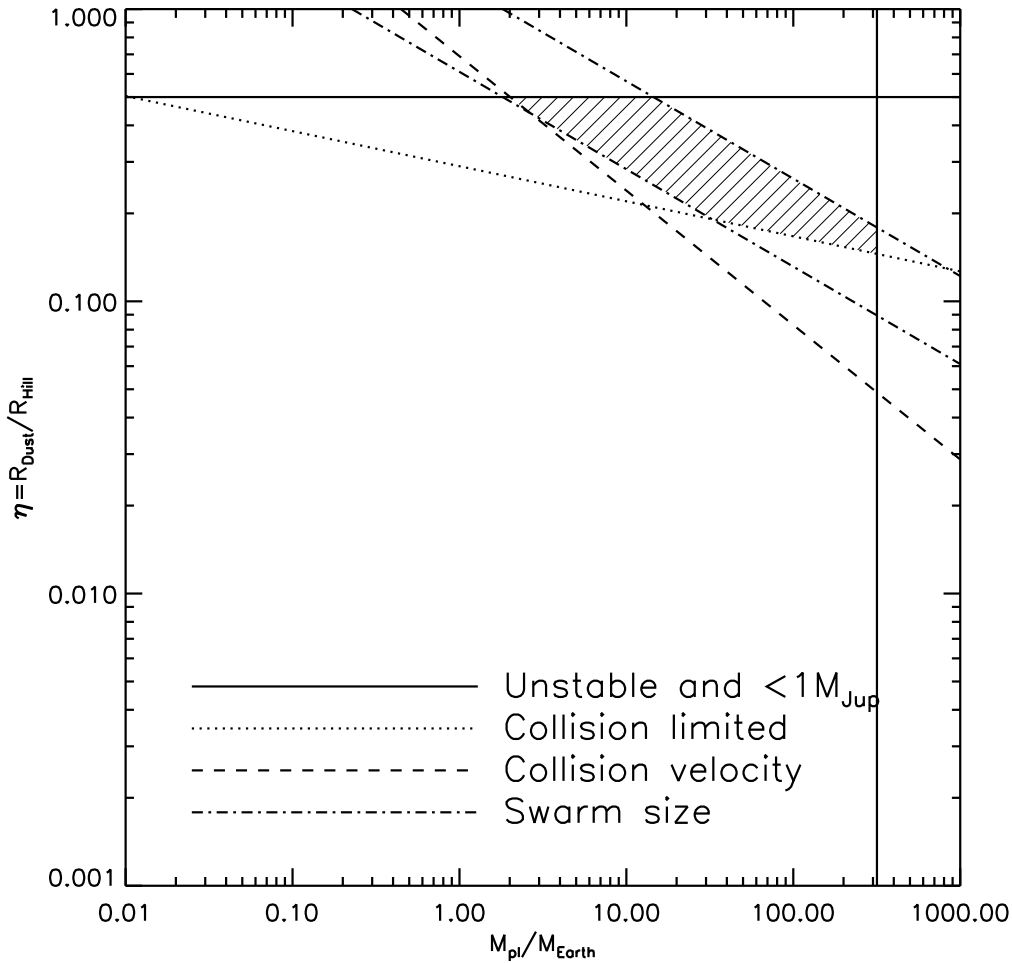}
\caption[]{\sl Semimajor axis~$\eta$ of satellites of the swarm versus planetary mass~$M_\mathrm{pl}$ diagram that shows the different constraints derived in the text. The parameters that could explain the Fomalhaut~b images are inside the dashed region.}
\label{fig : etavsmpl}
\end{figure*}

Moreover, KW11 assume that satellite orbits with~$\eta>0.5$ are not stable and do not consider them. Finally, we account for the~1\,M$_J$ upper limit that~\citet{janson12} put from the non detection at~$4.5\mu$m for a 400\,Myr system~\citep[close enough to~$440\pm40$\,Myr proposed by][]{mamajek12}. We plot all the constraints in Fig.\,\ref{fig : etavsmpl} that gives the semimajor axis~$\eta$ of the satellites against the planetary mass~$M_\mathrm{pl}$.
The parameters for which KW11's model can reproduce the photometry of a~1.16\,AU source are within the dashed area. The minimum and maximum planetary masses are~$\sim$2\,$M_\mathrm{Earth}$ and~1\,$M_J$ and the swarm has a total mass 2-11\,$M_\mathrm{Moon}$ and lies at 0.15-0.5 Hill radii around the planet.

 In the case of an unresolved object~(Fig.\,7 in~KW11), KW11 find that the mass of the planet ($<100\,M_\mathrm{Earth}$) is not sufficient for it to have a significant gaseous envelope and enable mechanisms that could explain the migration of Fomalhaut~b that presumably originates somewhere closer to the star. KW11 also argue that a single planet with mass $<100\,M_\mathrm{Earth}$ -- which is similar or less than the mass of the main debris ring~\citep[1-300\,$M_\mathrm{Earth}$,][]{wyatt02,chiang09} -- is unlikely responsible for shaping the dust belt. In our case of a source with diameter 1.16\,AU, the range of the planetary mass goes up to~$1\,M_J$ and a Jupiter-like planet can have a significant gaseous envelope and shape the dust belt.

\section{Conclusions}
Our independent analysis of the ACS, WFC3 and STIS data taken in 2004, 2006, 2009, and 2010 confirms that Fomalhaut~b is real and is not a speckle artifact as we clearly detect the object at the three epochs at several filters~(Figs.\,\ref{fig : im}, \ref{fig : im1}, and \ref{fig : im2}). In this way, we confirm the \citet[K08]{kalas08} detection.  However, we find differences in our analysis concerning astrometry and photometry of Fomalhaut~b.

Unlike~\citet{kalas10}, we cannot affirm that the object follows a trajectory that crosses the belt of dust because our astrometry is consistent within $1.16\,\sigma$ with crossing and non-crossing orbits~(\S\,\ref{subsec : astro}).

We detect Fomalhaut~b in the short wavelength filter F435W whereas K08 find an upper limit. We also derive an upper limit at F110W using WFC3. In the case of an unresolved source, our photometry is consistent with K08 at F606W/2004 and F814W/2006 but differs at F606W in the 2006 data~(\S\,\ref{subsubsec : photpsf}). As a consequence, unlike K08, we detect no significant variability of the F606W flux between 2004 and 2006. Considering the reduced and possible lack of variability at~F606W and the detection at F435W, several dust cloud models discussed by~K08 cannot be ruled out anymore~(\S\,\ref{subsubsec : cloudpsf}). K08 propose also a model of a Jovian planet surrounded by a large disk of dust. \citet{janson12} exclude this explanation mainly because of the variability at F606W and the assumed dust belt crossing trajectory. Given our new photometry and astrometry, we cannot reject this model~(\S\,\ref{subsubsec : diskpla}).

In the second part of our analysis, we study the possibility that Fomalhaut~b is spatially resolved in our images. The signal-to-noise ratios of the detections are low and more data are required to confirm the result but we find that our images are more consistent with an extended source with diameter 1.16\,AU than with a point source~(\S\,\ref{subsubsec : size} and \ref{subsubsec : instru}). The photometric variability of an extended source model at F606W is larger than for a point source but it is not yet significant~($<2.5\,\sigma$, \S\,\ref{subsubsec : photext}). Two models are considered to explain the size and the photometry of an extended source. First, the measurements are consistent with a cloud of dust produced by a collision of two Kuiper belt objects with radius 50\,km that would have occurred $\sim50-150$\,years ago~(\S\,\ref{subsubsec : kbocol}). The second model is an adaptation of the circumplanetary satellite swarm model proposed by~\citet{kennedy11}. It is consistent with the data when considering a 2\,$M_\mathrm{Earth}$-1\,$M_J$ planet surrounded by a swarm that lies at 0.15-0.5 Hill radii~(\S\,\ref{subsubsec : swarm}). 
 
The nature of the Fomalhaut~b object is still uncertain. However, from the two independent current and~K08 analysis of the~HST data, we can claim that Fomalhaut~b is a real object that orbits Fomalhaut~A.

\section{Acknowledgment}      
The authors are grateful to the~ACS team, John~Blakeslee, and Travis~Barman for helpful discussions. The authors also thank Paul~Kalas and James~Graham for useful communications on their analysis, and the anonymous referee for useful suggestions. Partial financial support for this research came from a~NASA grant to~UCLA.

\end{document}